\documentclass[10pt,a4papper]{article}
\usepackage[utf8x]{inputenc}
\usepackage[T2A]{fontenc}
\usepackage{indentfirst} 											
\usepackage[english]{babel}
\usepackage[]{graphicx}
\usepackage{geometry} 
\usepackage{float}

\begin{document}

\title{\textbf{Production of $\rho(770)$, $\eta$ pairs in decays $\rho(1450)\rightarrow \rho(770)\eta$, $\tau\rightarrow \rho(770)\eta\nu_{\tau}$, and in the process  $e^{+}e^{-} \rightarrow \rho(770)\eta$ in the Extended Nambu - Jona-Lasinio model}}
\author{M. K. Volkov$^{1}$\footnote{volkov@theor.jinr.ru}, K. Nurlan$^{1,2,3}$\footnote{nurlan.qanat@mail.ru},A . A. Pivovarov$^{1}$\footnote{tex$\_$k@mail.ru}\\
\small
\emph{$^{1}$ Joint Intsitute for Nuclear Research, Dubna, 141980, Russia}\\
\small
\emph{$^{2}$ Institute of Nuclear Physics, Almaty, 050032, Kazakhstan}\\
\small
\emph{$^{3}$Dubna State University, Dubna, 141980, Russia}}
\maketitle
\small

\begin{abstract}
Production of $\rho(770)$, $\eta$ pairs at colliding of electron-positron beams and in the decays $\rho(1450)\rightarrow \rho(770)\eta$, $\tau\rightarrow \rho(770)\eta\nu_{\tau}$ is calculated in the framework of the Nambu - Jona-Lasinio model. In the process of electron-positron annihilation and the $\tau$ - lepton decay the contributions of the intermediate $\rho(770)$ vector meson in the ground and first radially excited state are taken into account. The theoretical prediction for the decay $\tau\rightarrow \rho(770)\eta\nu_{\tau}$ is given. The obtained results for other processes are in satisfactory agreement with the experimental data.
\end{abstract}
\large

\section{Introduction}

Much attention is paid to the study of the interaction of mesons at low energies pays both in the experimental \cite{Delcourt:1982sj, Aubert:2007ym} and theoretical \cite{Volkov:2006kw, Volkov:2017arr} fields of physics. However, in theoretical calculations in the energy region below 2 GeV we can not apply the QCD perturbation theory and, therefore, it is necessary to use different phenomenological models  \cite{Volkov:2017arr, Volkov:2016umo}. One of the most popular and successfuly used phenomenological models of this type is the Nambu - Jona-Lasinio model (NJL)\cite{Volkov:2006kw, Ebert:1982pk, Volkov:1984kq, Volkov:1986zb, Ebert:1985kz, Vogl:1991qt, Klevansky:1992qe, Volkov:1993jw, Ebert:1994mf}, and in particular its extended modifications that allow one to describe in the framework of chiral symmetry the interactions not only of the ground states of four meson nonets (scalar, pseudoscalar, vector and axial-vector), but also their first radially excited states\cite{Volkov:2006kw, Volkov:2017arr, Volkov:2016umo, Volkov:1996br, Volkov:1996fk, Volkov:1997dd, Volkov:1999yi}. All calculations in the framework of the NJL model are carried out in the lowest order in $1/N_c$, where  $N_c$ is the number of colors of quarks. The allowance for higher degrees of decomposition can lead to a violation of chiral symmetry. Therefore, in the standard NJL model such decompositions are not used. However, there are other approaches to the formulation of phenomenological models at low energies, in particular, chiral perturbation theory  \cite{Gasser:1983yg, Gasser:1984ux}. As a result, it became possible to successfully describe many processes of meson production in colliding electron-positron beams and also in $\tau$ - lepton decays \cite{Volkov:2017arr}. 
In the present paper, we will continue these studies and describe processes $\rho'\rightarrow \rho\eta$, $\tau\rightarrow \rho\eta\nu_{\tau}$ и $e^{+}e^{-} \rightarrow \rho\eta$ in the framework of the extended NJL model. The theoretical results obtained here are in satisfactory agreement with the known experimental data \cite{Aubert:2007ym, Delcourt:1982sj, Donnachie:1991ep}.
 
\section{ Lagrangian of the extended NJL model for $\rho,\rho' ,\eta$  mesons } 

Lagrangian for vector $\rho,\rho'$  and pseudoscalar $\eta$ mesons in the extended NJL model takes the form \cite{Volkov:2017arr}:

\begin{equation}
\Delta L_{int}(q,\bar{q},\rho, \rho', \eta) = \bar{q}\left[ \frac{1}{2}\gamma^{\mu}\lambda_{\rho}(a_{\rho}\rho_{\mu} + b_{\rho}{\rho'}_{\mu})
+ i\gamma^{5}\lambda_{u} \sum_{\tilde{\eta} = \eta, \eta^{'}, \hat{\eta}, \hat{\eta}^{'}} A_{\tilde{\eta}}^{u} \tilde{\eta}\right]q,
\end{equation}

where $q$ and $\bar{q}$ are u and d quark fields with constituent masses $m_{u} = m_{d} = 280$ MeV,

\begin{displaymath}
a_{\rho} = \frac{1}{\sin(2\theta_{\rho}^{0})}\left[g_{\rho}\sin(\theta_{\rho} + \theta_{\rho}^{0}) +
g_{\rho'}f_{\rho}({\bf k}^{2})\sin(\theta_{\rho} - \theta_{\rho}^{0})\right],
\end{displaymath}
\begin{equation}
\label{Coefficients}
b_{\rho} = \frac{-1}{\sin(2\theta_{\rho}^{0})}\left[g_{\rho}\cos(\theta_{\rho} + \theta_{\rho}^{0}) +
g_{\rho'}f_{\rho}({\bf k}^{2})\cos(\theta_{\rho} - \theta_{\rho}^{0})\right],
\end{equation}
\begin{displaymath}
A_{\tilde{\eta}}^{u} = g_{1}^{u}b_{\tilde{\eta}1}^{u} + g_{2}^{u}f_{u}({\bf k}^{2})b_{\tilde{\eta}2}^{j},
\end{displaymath}
$f\left({\bf k}^{2}\right) = 1 + d_{uu} {\bf k}^{2}$ is the form factor for the descripttion of the first radially excited states, $d_{uu}$ is the slope parameter and $\theta_{\rho}$ and $\theta_{\rho}^{0}$ are the mixing angles for the mesons in the ground and excited states. The determination of the mixing angles can be found in \cite{Volkov:2017arr} (chapter 2.2)

\begin{displaymath}
d_{uu} = -1.784 \textrm{GeV}^{-2},
\end{displaymath}
\begin{equation}
\begin{array}{cccc}
  \theta_{\rho} = 81.8^{\circ},\\
 \theta_{\rho}^{0} = 61.5^{\circ}.
\end{array}
\end{equation}

The coefficients $b_{\tilde{\eta}1}^{j}, b_{\tilde{\eta}2}^{j}$ are the mixing parameters of the four $\eta$ meson states. These coefficients were obtained in the work \cite{Volkov:1999yi} (Chapter 3.3) and are given in Table 1. 

\begin{table}[h]
\caption{The mixing coefficients for the $\eta$ - mesons.}
\label{TabCoeff}
\begin{center}
\begin{tabular}{ccccc}
                        & $\eta$ & $\hat{\eta}$ & $\eta'$ & $\hat{\eta}'$ \\
$b_{\tilde{\eta}1}^{u}$ & 0.71   & 0.62         & -0.32   & 0.56          \\
$b_{\tilde{\eta}2}^{u}$ & 0.11   & -0.87        & -0.48   & -0.54         \\
$b_{\tilde{\eta}1}^{s}$ & 0.62   & 0.19         & 0.56    & -0.67         \\
$b_{\tilde{\eta}2}^{s}$ & 0.06   & -0.66        & 0.30    & 0.82
\end{tabular}
\end{center}
\end{table}

The coupling constsnts are:
\begin{displaymath}
g_{\rho} = \left(\frac{2}{3}I_{2}(m_{u},m_{u})\right)^{-1/2} \approx 6.14,
\quad g_{\rho'} = \left(\frac{2}{3}I_{2}^{f_{uu}^{2}}(m_{u},m_{u})\right)^{-1/2} \approx 9.87,
\end{displaymath}
\begin{displaymath}
g_{1}^{u} = \left(\frac{4}{Z_{\pi}}I_{2}(m_{u},m_{u})\right)^{-1/2} \approx 3.02,
\quad g_{2}^{u} = \left(4I_{2}^{f_{uu}^{2}}(m_{u},m_{u})\right)^{-1/2} \approx 4.03,
\end{displaymath}
where $Z_{\pi}$ - is the factor corresponding to the $\pi-a_1$ transitions.  
\begin{displaymath}
Z_{\pi} = \left(1 - 6\frac{m^{2}_{u}}{M^{2}_{a_{1}}}\right)^{-1} \approx 1.45, 
\end{displaymath}
where $M_{a_1}=1230MeV$ is the mass of the axial-vector $a_1$ meson, integral $I_{2}$ has the following form:

\begin{equation}
I_{2}^{f^{n}}(m_{1}, m_{2}) =
-i\frac{N_{c}}{(2\pi)^{4}}\int\frac{f^{n}({\bf k}^{2})}{(m_{1}^{2} - k^2)(m_{2}^{2} - k^2)}\theta(\Lambda_{3}^{2} - {\bf k}^2)
\mathrm{d}^{4}k,
\end{equation}

$\Lambda_{3} = 1.03$ GeV is the cutoff parameter \cite{Ebert:1993xi}.  

We also give the Lagrangian of the weak interaction of the leptonic current with quarks, which is necessary for describing the decay $\tau\rightarrow \rho\eta\nu_{\tau}$.  

\begin{equation}
 L^{weak}=\bar{\tau}\gamma_{\mu}(1-\gamma_{5})\nu \frac{G_{F}}{\sqrt{2}}V_{ud}\bar{d}(1-\gamma_{5})\gamma_{\mu}u,
\end{equation}

where $G_{F}=1.1663787(6)\times{10}^{-11}{MeV}^{-2}$ is the Fermi constant, ${V}_{ud}$ is the element of the Cabibbo-Kobayashi-Maskawa matrix.

\section{The decay $\rho'\rightarrow \rho\eta$}

The process $\rho'\rightarrow \rho\eta$ in the one-loop quark approximation is described by the diagram in Fig.1. The corresponding amplitude has the form:

\begin{equation}
T = 4m_{u}I^{\rho'\rho\eta}_{03} \epsilon^{\nu\lambda\delta\sigma} e_{\nu}(l) e_{\lambda}(p)p^{\delta}_\rho q^{\sigma}_\eta  ,
\end{equation}
where  $e_{\nu}(l) e_{\lambda}(p)$ are the polarization vectors of $\rho'$, $\rho$ mesons. The loop integral $I^{\rho'\rho\eta}_{03}$ has the form: 
 
 \begin{equation}
I^{\rho'\rho\eta}_{03}(m_{u}, m_{d}) =
-i\frac{N_{c}}{(2\pi)^{4}}\int\frac{a_{\rho}b_{\rho}A_{\eta}^{u} }{(m_{u}^{2} - (k+p)^{2})(m_{u}^{2} - {k}^{2})(m_{u}^{2} - (k-q)^{2})}\theta(\Lambda_{3}^{2} - {\bf k}^2)
\mathrm{d}^{4}k,
\end{equation}
 $a_{\rho}$, $b_{\rho}$,$A_{\eta}^{u}$ are the coefficients containing mixing angles, coefficients and form factors. 
\begin{figure}[h] 
\center{\includegraphics[width=0.2\linewidth]{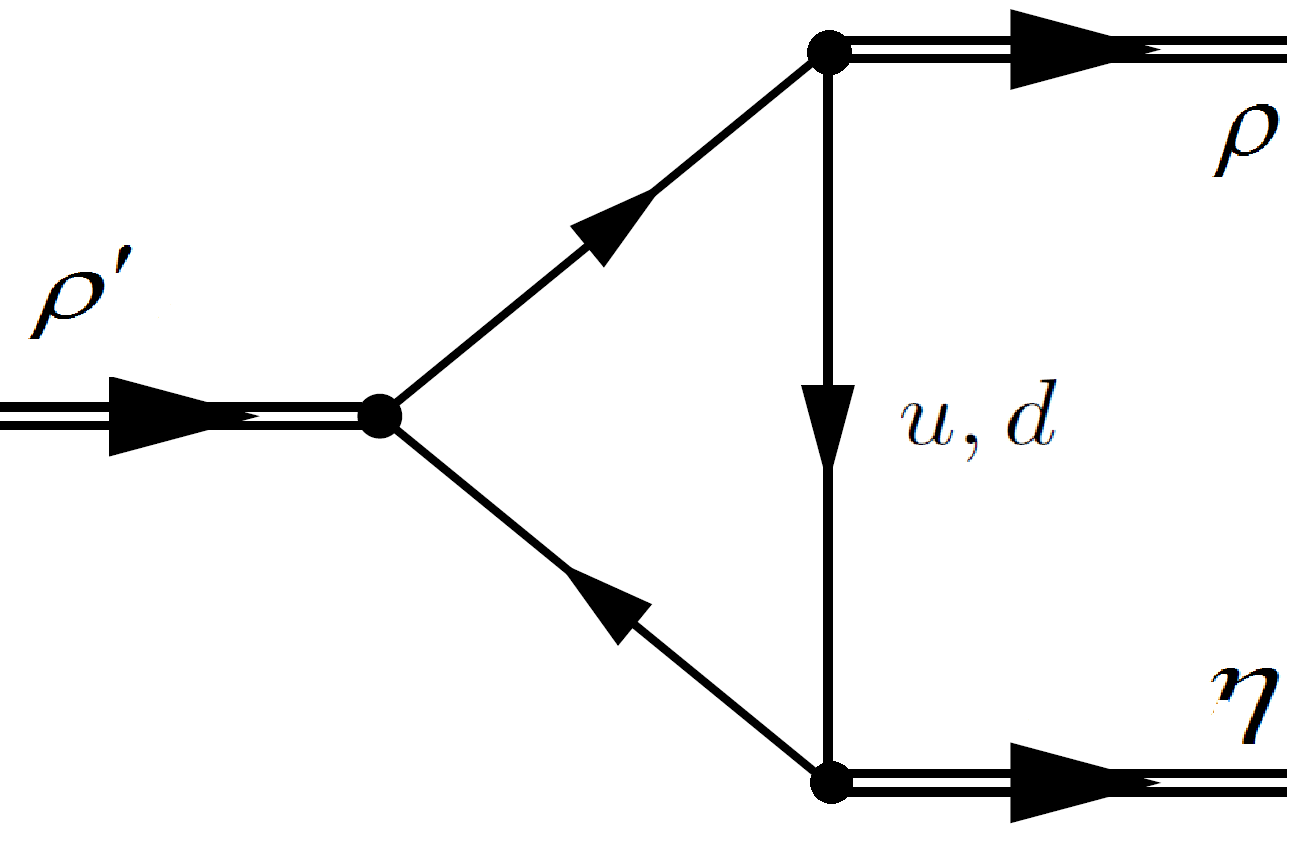}} 
\linespread{1} \caption {The diagram describing the decay $\rho'\rightarrow \rho\eta$ } 
\label{pic:pic1} 
\end{figure}
 
As a result, for the width of the decay we have:
\begin{equation}
\Gamma_{NJL}(\rho'\rightarrow \rho\eta)=10.8 MeV.
\end{equation}

Similar values ​​for our results for the width of the decay $\rho'\rightarrow \rho\eta$ are obtained in the paper \cite{Piotrowska:2017rqt}. Unfortunately, the experiment gives only a limitation from above for this decay \cite{Donnachie:1991ep}:

\begin{equation}
\Gamma_{exp}(\rho'\rightarrow \rho\eta) < 16.0\pm2.4 MeV.
\end{equation}
  
\section{The process $e^{+}e^{-} \rightarrow \rho\eta$ in the extended NJL model }
  
The diagrams describing the process $e^{+}e^{-} \rightarrow \rho\eta$ are shown in Figures 2 and 3. The corresponding amplitude is:

\begin{equation}
T = \frac{4 m_{u}e^2}{s} l^{\mu} \left\{B_{(\gamma)}+B_{(\rho)}+B_{(\rho')} \right\}_{\mu\nu} \epsilon^{\nu\lambda\delta\sigma} e_{\lambda}(p)p^{\delta}_\rho q^{\sigma}_\eta,
\end{equation}
 where  $l^{\mu} = \bar{e}\gamma^{\mu}e$ is the lepton current, $s = (p(e^{-}) + p(e^{+}))^2$.

 The contribution of the contact diagram (Fig. 2) gives: 
\begin{equation}
{B_{(\gamma)}}_{\mu\nu} =I^{\gamma\rho\eta}_{03}g_{\mu\nu}.   
\end{equation}

The contribution of the diagrams with intermediate $\rho$, $\rho'$ mesons: 
\begin{equation}
{B_{(\rho)}}_{\mu\nu} = \frac{C_{\rho}}{g_{\rho}}I^{\rho\rho\eta}_{03}\frac{g_{\mu\nu}s-p_{\mu}p_{\nu}}{M_{\rho}^{2}-s-i\sqrt{s}\Gamma_{\rho}},
\end{equation}
\begin{equation}
{B_{(\rho')}}_{\mu\nu} = \frac{C_{\rho'}}{g_{\rho}}I^{\rho'\rho\eta}_{03}\frac{g_{\mu\nu}s-p_{\mu}p_{\nu}}{M_{\rho'}^{2}-s-i\sqrt{s}\Gamma_{\rho'}}.
\end{equation}

$M_{\rho} = 775$ MeV, $M_{{\rho'}} = 1465$ MeV,
$\Gamma_{\rho} = 149$ MeV, $\Gamma_{\rho'} = 400$ MeV are the masses of vector mesons and their full widths \cite{Agashe:2016kda}. 
The constant $C_{\rho}$ describes the transition of a virtual photon into an intermediate vector meson: 
\begin{equation}
C_{\rho} = \frac{1}{\sin\left(2\theta_{\rho}^{0}\right)}\left[\sin\left(\theta_{\rho} + \theta_{\rho}^{0}\right) +
R_{V}\sin\left(\theta_{\rho} - \theta_{\rho}^{0}\right)\right],
\end{equation}

\begin{displaymath}
R_{V} = \frac{I_{2}^{f}(m_{u},m_{d})}{\sqrt{I_{2}(m_{u},m_{d})I_{2}^{f^{2}}(m_{u},m_{d})}}.
\end{displaymath}

\begin{figure}[H] 
\center{\includegraphics[width=0.3\linewidth]{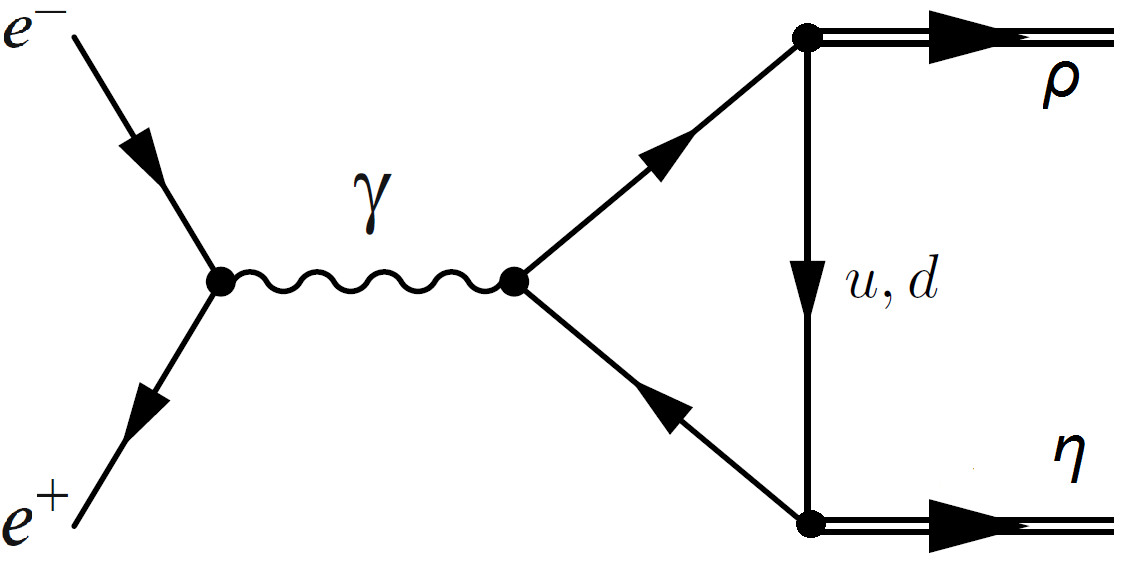}} 
\linespread{1} \caption {The contact diagram of the process $e^{+}e^{-} \rightarrow \rho\eta$.  } 
\label{pic2:pic2} 
\end{figure}

\begin{figure}[H] 
\center{\includegraphics[width=0.5\linewidth]{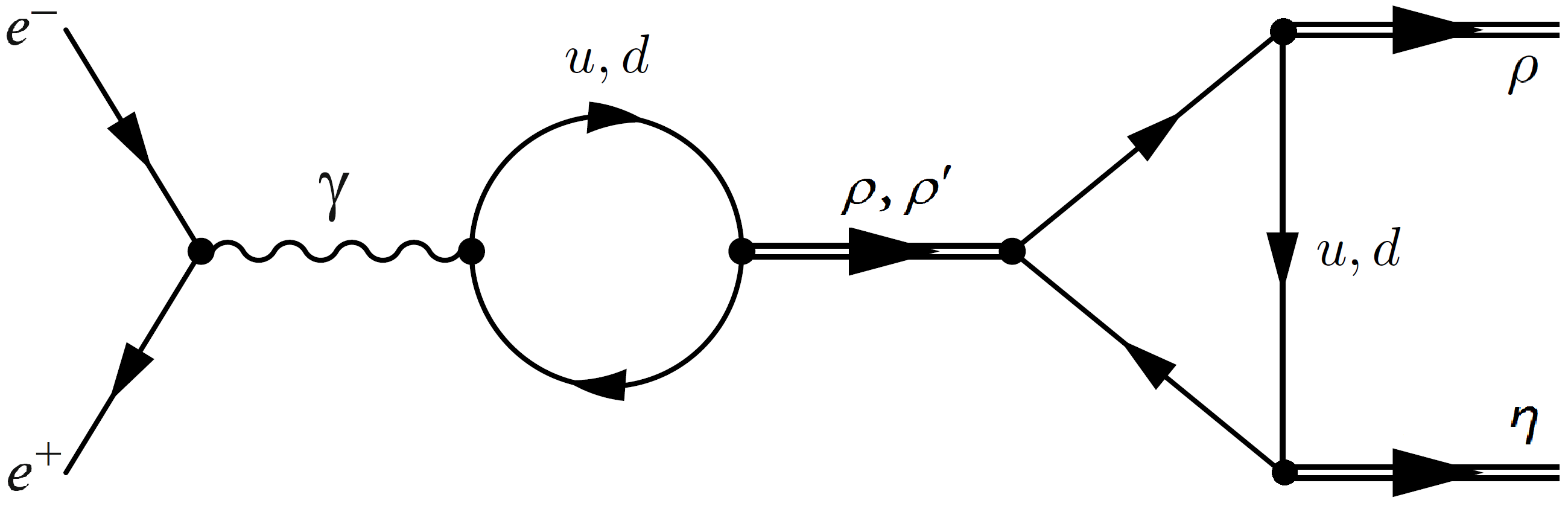}} 
\linespread{1} \caption { The diagram of the process $e^{+}e^{-} \rightarrow \rho\eta$ with intermediate $\rho$,  $\rho'$ mesons. } 
\label{pic3:pic3} 
\end{figure}

Numerical results of calculations are given in Fig. 4. The theoretical result corresponds to the solid line and the experimental data \cite{Delcourt:1982sj} are shown as single points.

\begin{figure}[H]
\center{\includegraphics[width=0.5\linewidth]{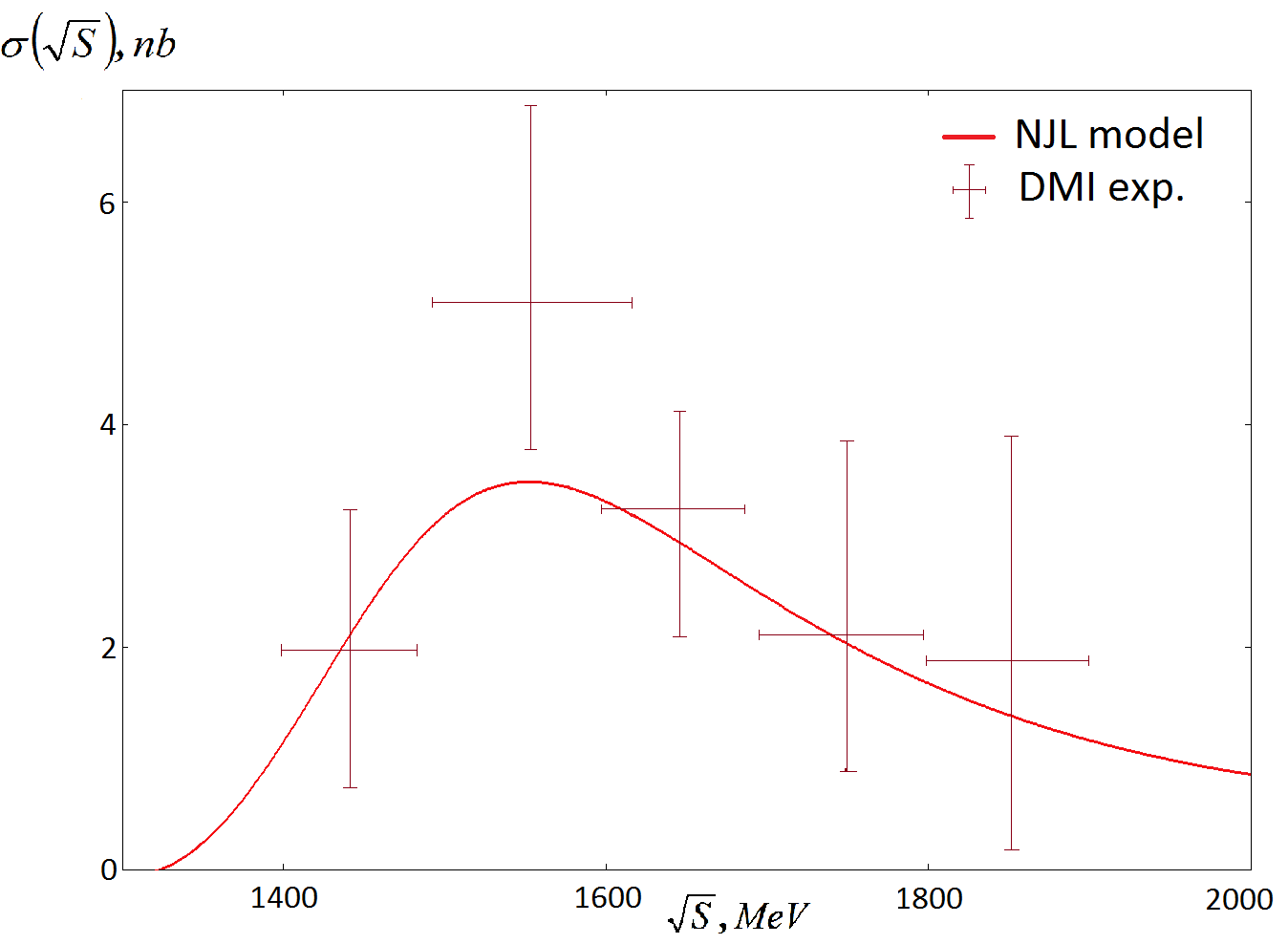}} 
\linespread{1} \caption {Comparison of the predictions of the NJL model for the process $e^{+}e^{-} \rightarrow \rho\eta$ with the experimental data \cite{Delcourt:1982sj}. } 
\label{pic3:pic4} 
\end{figure}

\section{The decay $\tau\rightarrow \rho\eta\nu_{\tau}$ }
The diagrams of the decay $\tau\rightarrow \rho\eta\nu_{\tau}$ are shown in Figures 5 and 6. The amplitude of the decay is close to that described in Section 4 and has the form: 

\begin{displaymath}
T = -i4m_{u}\frac{G_{F}}{\sqrt{2}} l^{\mu} V_{ud} \left\{I^{\gamma\rho\eta}_{03}g_{\mu \nu}+\frac{C_{\rho}}{g_{\rho}}I^{\rho \rho \eta}_{03}\frac{g_{\mu \nu}s-p_{\mu}p_{\nu}}{M_{\rho}^{2}-s-i\sqrt{s}\Gamma_{\rho}} \right.
\end{displaymath}
\begin{equation}
\left. + \frac{C_{\rho'}}{g_{\rho}}I^{\rho' \rho \eta}_{03} \frac{g_{\mu \nu}s-p_{\mu}p_{\nu}}{M_{\rho'}^{2}-s-i\sqrt{s}\Gamma_{\rho^{'}}}\right\}\epsilon^{\nu \lambda \delta \sigma}e_{\lambda}(p)p^{\delta}_{\rho} q^{\sigma}_{\eta}.
\end{equation}

Here, instead of an intermediate photon, there is a W-boson, the role of neutral intermediate vector $\rho$ mesons will be played by charged vector mesons.

\begin{figure}[H]
\center{\includegraphics[width=0.35\linewidth]{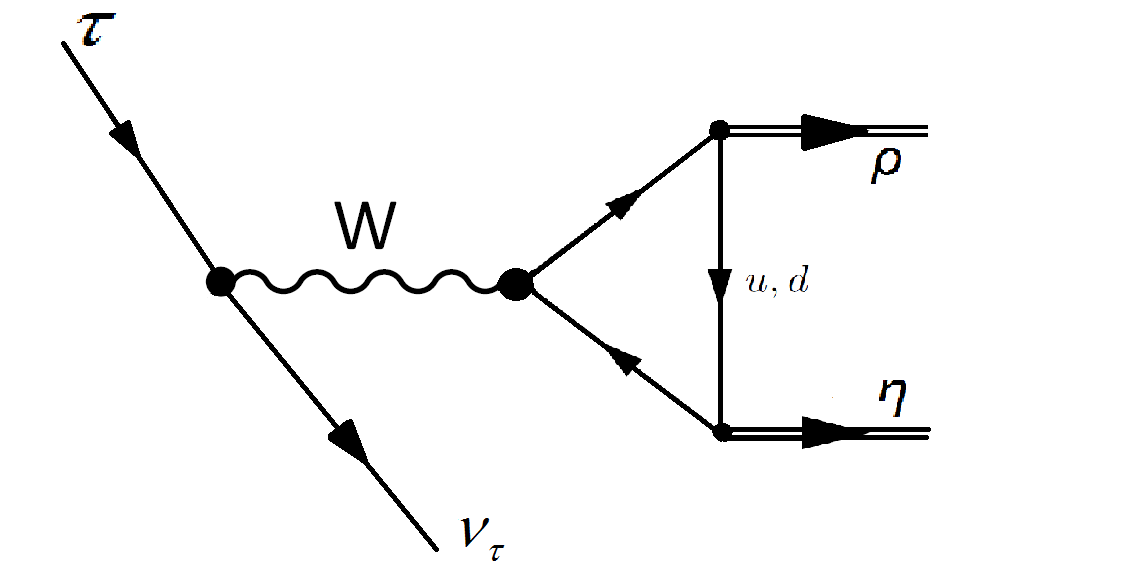}} 
\linespread{1} \caption {The contact diagram of the decay $\tau\rightarrow \rho\eta\nu_{\tau}$. } 
\label{pic3:pic5} 
\end{figure}

\begin{figure}[H]
\center{\includegraphics[width=0.5\linewidth]{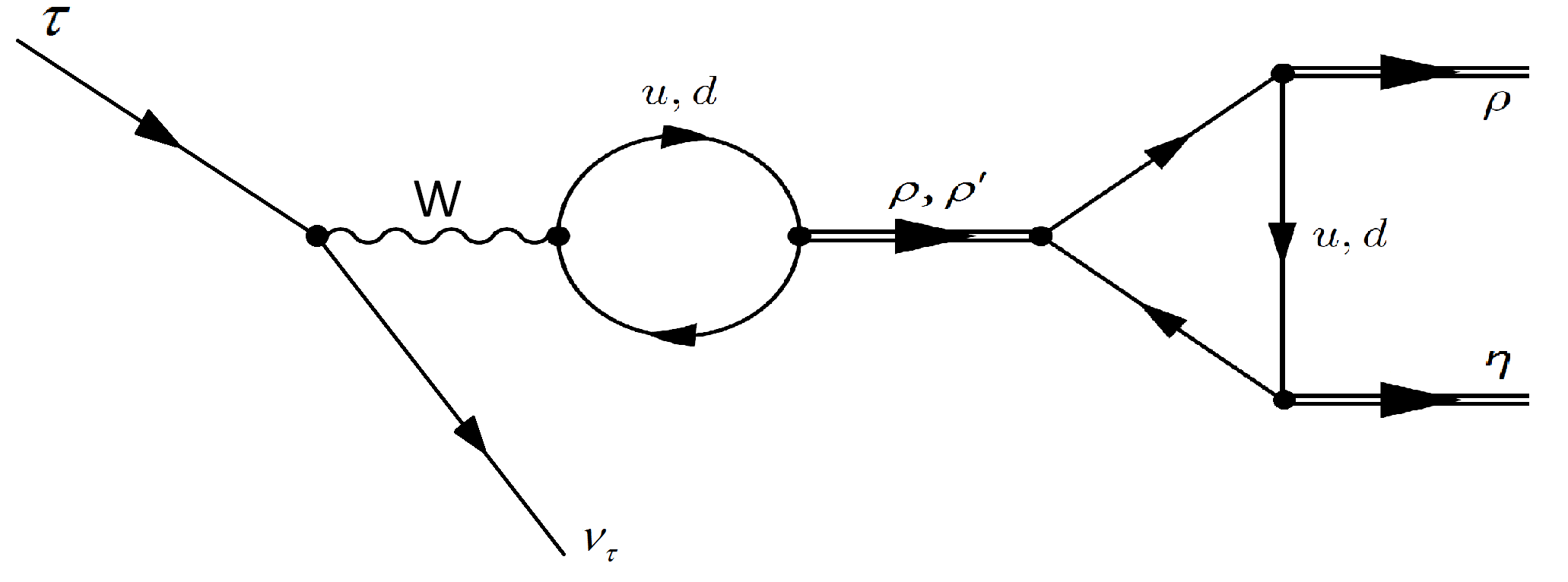}} 
\linespread{1} \caption { The diagram of the decay $\tau\rightarrow \rho\eta\nu_{\tau}$ with intermediate $\rho$, $\rho'$ mesons. } 
\label{pic3:pic6} 
\end{figure}

As a result, for the partial width of the decay $\tau\rightarrow \rho\eta\nu_{\tau}$ we obtain:

\begin{equation}
Br(\tau\rightarrow \rho\eta\nu_{\tau})_{NJL}=1.44\times10^{-3}.
\end{equation}

\section{Conclusion}
The theoretical results obtained for $\rho'\rightarrow \rho\eta$ and $e^{+}e^{-} \rightarrow \rho\eta$ are in satisfactory agreement with the known experimental data \cite{Delcourt:1982sj, Donnachie:1991ep}. Our theoretical estimates for the width of the decay $\rho'\rightarrow \rho\eta$ in the framework of the extended NJL model are close to the theoretical result obtained in another phenomenological model\cite{Piotrowska:2017rqt}.Unfortunately, for the decay of $\tau\rightarrow \rho\eta\nu_{\tau}$, there are currently no direct experimental data, but the width of the decay $\tau\rightarrow \rho\eta\nu_{\tau}$ was fairly well measured and  is equal to ${Br(\tau\rightarrow 2\pi\eta\nu_{\tau})}_{exp}=(1.39\pm0.1)\times10^{-3}$  \cite{Agashe:2016kda}. We note that in order of magnitude this value is close to the value of the decay width $\tau\rightarrow \rho\eta\nu_{\tau}$. This allows us to hope that our prediction for the width $\tau\rightarrow \rho\eta\nu_{\tau}$ is within reasonable limits in comparison with the width $\tau\rightarrow 2\pi\eta\nu_{\tau}$ and in the near future will find experimental confirmation.

In conclusion, we note that the width of the process $\tau\rightarrow 2\pi\eta\nu_{\tau}$ was previously calculated within the extended NJL model in satisfactory agreement with the experimental data \cite{Arbuzov:2013zba}.

\section*{Acknowledgments}
We are greatful to A.B. Arbuzov for useful disscussions.

\end{document}